\documentclass[twocolumn,showpacs,amsmath,amssymb,aps,showkeys,floatfix,prd,a4paper]{revtex4}

\usepackage[dvips]{graphicx}
\usepackage{dcolumn}
\usepackage{bm}
\usepackage{epsfig}
\usepackage{amsmath}
\usepackage{amsfonts}
\usepackage{amssymb,amscd}
\usepackage[latin1]{inputenc}
\usepackage{units}
\usepackage{hyperref}

\begin{document}


\title{Diffractive photon production at high momentum transfer in $ep$ collisions}

\author{V.~P. Gon\c{c}alves}
\email{barros@ufpel.edu.br}
\author{W.~K. Sauter}
\email{werner.sauter@ufpel.edu.br}
\affiliation{High and Medium Energy Group, \\
Instituto de F\'{\i}sica e Matem\'atica, Universidade Federal de Pelotas\\
Caixa Postal 354, CEP 96010-900, Pelotas, RS, Brazil}
\date{\today}

\begin{abstract}
The diffractive photon production at large momentum transfer and large energies is a probe of the parton dynamics of the diffractive exchange. In this paper we revisit the leading order (LO) BFKL equation approach for this process and estimate, for the first time, the differential and  total cross sections considering the  
next-to-leading order (NLO) corrections for the BFKL characteristic function.  We obtain  a reasonable agreement with the DESY HERA data.
\end{abstract}


\maketitle

\section{Introduction \label{sec:intro}}

The description of exclusive diffractive processes has been proposed as a probe of the Quantum Chromodynamics (QCD) dynamics in the high energy limit (For a recent reviews see, e.g. Ref. \cite{Schoeffel:2009aa}). It is expected that the study of these processes provide insight into the parton dynamics of the diffractive exchange  when a hard scale is present. In particular, the diffractive vector meson and photon production at large momentum transfer is expected to probe the QCD Pomeron, which is described by the Balitsky, Fadin, Kuraev, and Lipatov (BFKL) equation \cite{Lipatov:1976zz,Kuraev:1976ge,Kuraev:1977fs,Balitsky:1978ic}.
In the last years, the H1 and ZEUS collaborations at DESY-HERA have measured the exclusive production of $\rho$, $\phi$, $J/\Psi$ and $\gamma$ with hadron dissociation and  large values of $|t|$,  the   square of the four-momentum  transferred across the associated rapidity gap.  The experimental data for vector meson production are quite well described in terms of the BFKL formalism at leading order \cite{Forshaw:2001pf,Enberg:2002zy,Enberg:2003jw,Poludniowski:2003yk,Goncalves:2009gs,Goncalves:2010zb}. In contrast,  for  photon production, the  analysis presented in the  Ref. \cite{Aaron:2008cq} indicate 
that  this approach describes the energy dependence of the cross section but it is unable to describe its $|t|$ dependence. 
It is important to emphasize that in \cite{Aaron:2008cq} the free parameters present in the BFKL formalism at leading order has been constrained using the experimental data for the total cross section and the differential cross section was predicted and compared with the data.
This procedure is the opposed from that used in Refs. \cite{Goncalves:2009gs,Goncalves:2010zb} in order to describe the vector meson data, where the experimental data for the differential cross section for a fixed energy were used to constrain the free parameters and the energy dependence of the total cross section was predicted by the LO BFKL equation. The possibility that the use of a different procedure allow us to obtain a better description of the experimental data is our first motivation to revisit the description of  the diffractive photon production in the BFKL formalism. The second one is associated to the fact that in the last years the next-to-leading order (NLO) corrections to the BFKL kernel were determined \cite{Fadin:1998py,Ciafaloni:1998gs}  and the origin of the instabilities in the perturbative series was understand  and the problem was solved using methods based on the combination of collinear and small-$x$ resummations \cite{Salam:1998tj,Brodsky:1998kn,Khoze:2004hx, Vera:2005jt}. It allow us to include, for the first time,  the NLO corrections for the BFKL characteristic function in the analysis of the diffractive photon production at large-$|t|$. This is a first step in the direction of a full NLO calculation, since we are not including the NLO corrections to the impact factor associated to the photon-photon transition. It has been calculated in Refs.  \cite{Fadin:2002tu,Bartels:2000gt,Bartels:2001mv,Bartels:2002uz,Bartels:2004bi} and its analytical form was recently presented in  \cite{Balitsky:2012bs}. Thus, we are aware that our phenomenological study  is a educated guess to a complete NLO calculation. However, we hope that this gives some enlightenments about the underlying QCD dynamics.

The paper is organized as follows. In the next Section we present a brief  discussion about  the diffractive photon production at large-$t$ in $ep$ collisions. In Section  \ref{sec:bfkl} we present the formalism for the calculation of the cross section is presented and the next-to-leading order corrections to the BFKL kernel are briefly revised, as well as, the different schemes used to improve the convergence of the perturbative series. In Section \ref{sec:res} we present our predictions and compare with the available data. Finally, a summary of our main conclusions is presented in  Section \ref{sec:concl}.


\section{Diffractive photon production  \label{sec:photon}}

Diffractive processes such as $ep \rightarrow e XY$, where $X$ and $Y$ are hadronic systems of the dissociated photon and proton, respectively, have been studied extensively at HERA in the last decades (For a recent review see, e.g., Ref. \cite{Schoeffel:2009aa}). 
One of the cleanest of the diffractive processes is that of diffractive photon production ($X = \gamma$), with the final state photon  having a large transverse momentum and being well separated in rapidity from the hadronic system $Y$.  
The cross section for the  process of electron - proton scattering measured in HERA is directly related with photon-proton scattering by the relation~\cite{Aaron:2008cq},
\begin{equation}
\frac{d\sigma(ep \rightarrow e+\gamma +Y)}{dQ^2 dy dt} = \Gamma(y,Q^2) \frac{d\sigma (\gamma^\ast p \rightarrow \gamma + Y)}{dt}
\end{equation}
where $y = W^2/s$, $W$ is the center-of-mass energy of the $\gamma^\ast p$ system and    $\Gamma(y,Q^2)$ is the Weizs\"acker-Williams approximation of the flux of photons produced by the electron   given by \cite{Frixione:1993yw}
\begin{equation}
\Gamma  (y,Q^2) = \frac{\alpha_\mathrm{em}}{2\pi} \left[\frac{2m^2_ey}{Q^4} + \frac{1+(1-y)^2}{yQ^2}\right]\,\,.
\end{equation}
 In the photoproduction regime studied by the H1 Collaboration, this flux is integrated over the range $Q^2 < Q^2_\mathrm{max} = \unit[0.01]{GeV^2}$ ~\cite{Aaron:2008cq}, which implies 
\begin{equation}
\Gamma(y) = \frac{\alpha_\mathrm{em}}{2\pi} \left[ 2m^2_ey\left(\frac{1}{Q^2_\mathrm{max}} \right) + \frac{1+ (1-y)^2}{y} \ln\frac{Q^2_\mathrm{max}}{Q^2_\mathrm{min}}\right],
\end{equation}
with  $Q^2_\mathrm{min} = m_e^2y^2/(1-y)$.

\begin{figure}[t]
\begin{center}
\includegraphics*[width=8cm]{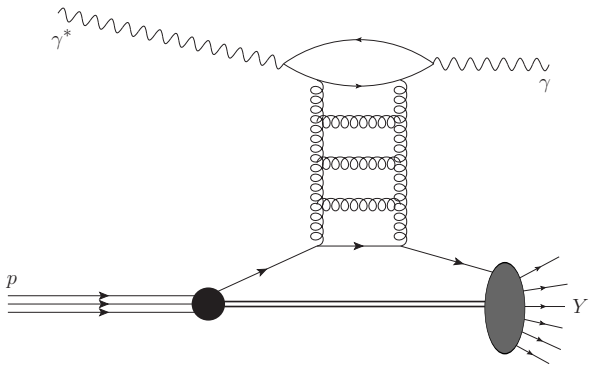} 
\end{center}
\caption{Diffractive production of a photon with high-$t$ exchange (hard Pomeron) in a virtual photon-proton scattering, producing a hadronic ($Y$) system.}
\label{fig:proc}
\end{figure} 

Although 
parton fragmentation, production of a resolved photon and the Bethe-Heitler process contribute for this process \cite{Hoyer:2000mb},  the exchange of gluons in the $t$ channel dominates over the other contributions at large $s$, which implies that  the leading contribution in high energy regime comes from the Pomeron exchange (See Fig.    \ref{fig:proc}).  In this regime, the cross section for the process photon-proton can be fully calculated in the factor impact representation  and is given by the convolution of the hard partonic cross-section with the parton distribution function in the proton \cite{Ginzburg:1985tp,Ginzburg:1996vq,Ivanov:1998jw,Cox:1999kv,Evanson:1999zb}. The final expression is
\begin{eqnarray}
\frac{d\sigma (\gamma^\ast p \rightarrow \gamma +Y)}{dt dx_j} = \left[ \frac{81}{16} G(x_j,|t|) + \nonumber \right.  \\ \left. \sum_j ( q_j(x_j,|t|) + \bar{q}_j(x_j,|t|))\right] \, \frac{d\hat{\sigma}}{dt}(\gamma^\ast q \rightarrow \gamma q)\,,
\label{eq:dsigdtdx}
\end{eqnarray}
where $G(x_j,|t|)$ and $q_j(x_j,|t|)$ are the proton gluon and quark distribution functions, respectively, and ${d\hat{\sigma}}/
{dt}$ is the differential cross section for the subprocess $\gamma^\ast q \rightarrow \gamma q$. The basic idea is that at large - $|t|$, the gluonic ladder probes a parton inside the target and the struck parton initiates a jet which carries the fraction $x_j$ of the longitudinal momentum of the incoming hadron, being given by $x_j = -t/(-t + M_Y^2 - m^2)$, where $M_Y$ is the mass of the {products of the target dissociation} and $m$ is the mass of the target. The minimum value of $x_j$, $x_\mathrm{min}$, is calculated considering the experimental cuts on $M_Y$. In what follows we consider  a fixed value $x_\mathrm{min} = 10^{-2}$  following Ref.  \cite{Aaron:2008cq} and the theoretical studies presented in Refs. \cite{Forshaw:2001pf,Enberg:2002zy,Enberg:2003jw,Poludniowski:2003yk,Goncalves:2009gs,Goncalves:2010zb}. Experimentally, in diffractive photon production at large - $|t|$ a backscattering photon is produced with a small angle in the detector and the transverse momentum of the photon is counterbalanced by the  momentum of the jet  scattered in the frontal region. The photon and the jet are separated by a large rapidity gap among them.

The diffractive photon production can be modelled in the rest frame of the proton by a sequence of three events:  first, the fluctuation of the incoming photon into a quark-antiquark pair a long distance from the proton target. Second, the pair interacts with the parton $q$ via the exchange of a color singlet state (the gluon ladder). Finally, the final $q\bar{q}$ pair annihilates into a real photon.  In the interaction of the ladder  with the parton in the proton,  it transfers  momentum and dissociates  the proton, whose fragments, aftermost, hadronize. This process has been studied in Refs.  \cite{Ginzburg:1985tp,Ginzburg:1996vq,Ivanov:1998jw,Cox:1999kv,Evanson:1999zb} and its description is closely related to the diffractive production of vector mesons at large $|t|$ discussed in detail in Refs. \cite{Forshaw:2001pf,Enberg:2002zy,Enberg:2003jw,Poludniowski:2003yk}. The main advantage present in photon production is that the theoretical calculations are simplified by the absence of a vector meson wave function, with the only non-perturbative part being the parton distribution functions of the proton. However, the cross section is suppressed relative to that of vector meson production by the electromagnetic coupling of the $q\bar{q}$ pair to the final state photon.  In the next section we  present a brief review of the description of this process in the BFKL formalism.

\section{BFKL formalism  \label{sec:bfkl}}

The energy and momentum transfer dependences of the cross section for the diffractive photon production is directly related to the description of the differential cross section for the partonic process $\gamma^\ast q \rightarrow \gamma q$, which can be expressed by \cite{Ivanov:1998jw,Evanson:1999zb},
\begin{equation}
\frac{d\hat{\sigma}}{dt}(\gamma^\ast q \rightarrow \gamma q) = \frac{1}{16\pi s^2}\left\{ |{\mathcal{A}}_{(+,+)}(s,t)|^2 + |{\mathcal{A}}_{(+,-)}(s,t)|^2\right\}\,\,,\label{eq:dsdtp}
\end{equation}
where the scattering amplitudes $\mathcal{A}_{(+,\pm)}$ depend of the polarization states of the photons (represented by the plus/minus signs) through the impact factors for the transition $\gamma^\ast \rightarrow \gamma$. As the Pomeron couples directly to the partons in the proton, the scattering amplitude factorizes in the impact factor associated to the transition of the virtual photon into the real photon, the gluonic ladder and the coupling of the Pomeron with the parton in the proton.  At leading order the amplitudes  are given by \cite{Evanson:1999zb} 
\begin{widetext}
\begin{eqnarray}
{\mathcal{A}}_{(+,+)} &=& i \alpha \alpha_s^2 \sum_q e_q^2 \frac{\pi}{6} \frac{s}{|t|} \int\!d\nu \frac{\nu^2}{(1/4 + \nu^2)^2} \frac{\tanh(\pi\nu)}{\pi\nu(1+\nu^2)} \left[ \frac{s}{s_0} \right]^{\omega(\nu)} \int^{1/2+i\infty}_{1/2-i\infty} \frac{dz}{2\pi i}\left[ \frac{Q^2}{|t|} \right]^{-z/2}  \nonumber \\ 
 &\times & \frac{\Gamma(1/2 -i\nu -z/2) \Gamma(1/2 +i\nu -z/2)}{|\Gamma(1/2+i\nu)|^2} \Gamma(z/2)\Gamma(z/2+1) \left[ z^2+ 11 + 12\nu^2 \right] \\
{\mathcal{A}}_{(+,-)} &=& i \alpha \alpha_s^2 \sum_q e_q^2 \frac{2\pi}{3} \frac{s}{|t|} \int\!d\nu \frac{\nu^2}{(1/4 + \nu^2)^2} \frac{\tanh(\pi\nu)}{\pi\nu(1+\nu^2)} \left[ \frac{s}{s_0} \right]^{\omega(\nu)} \int^{1/2+i\infty}_{1/2-i\infty} \frac{dz}{2\pi i}\left[ \frac{Q^2}{|t|} \right]^{-z/2}  \nonumber \\ 
 &\times & \frac{\Gamma(3/2 -i\nu -z/2) \Gamma(3/2 +i\nu -z/2)}{|\Gamma(1/2+i\nu)|^2} \Gamma(z/2)\Gamma(z/2+1) 
\label{amplitudes}\end{eqnarray}
\end{widetext}
where $\alpha$ is fine structure constant; $\alpha_s$ is the strong coupling constant (comes from the impact factor and will be kept fixed); $e_q$ is the quark charges; $s_0$ is a energy scale, chosen to be $c_1 \cdot Q^2 + c_2 \cdot|t|$ ($c_1$ and $c_2$ are free parameters) as the meson production case \cite{Goncalves:2009gs} (see discussion below); $\omega(\nu)$ is the BFKL characteristic function; $Q^2$ is the photon virtuality and $\Gamma(z)$ is the gamma function.
These results have been derived using the  Mueller-Tang~\cite{Mueller:1992pe} prescription for the parton - Pomeron coupling and  only have considered the contribution associated to the lowest conformal spin ($m=0$).  In  Ref.  \cite{Munier:1999jm} these results have been rederived and it was shown that the only one non-zero contribution for highest conformal spin, $m=2$, is small.   In the photoproduction and large-$t$  regimes of interest in our analysis ($|t| \ge 4$  GeV$^ 2$) the amplitudes are given by (For details see  \cite{Ivanov:1998jw})
\begin{widetext}
\begin{eqnarray}
{\mathcal{A}}_{(+,+)} &=& i \alpha \alpha_s^2 \sum_q e_q^2 \frac{4\pi}{3} \frac{s}{|t|} \int\!d\nu \frac{\nu^2}{(1/4 + \nu^2)^2} \frac{\tanh(\pi\nu)}{\pi\nu} \frac{11/4+3\nu^2}{1+\nu^2} \left[ \frac{s}{s_0} \right]^{\omega(\nu)}   \\
{\mathcal{A}}_{(+,-)} &=& i \alpha \alpha_s^2 \sum_q e_q^2 \frac{4\pi}{3} \frac{s}{|t|} \int\!d\nu \frac{\nu^2}{(1/4 + \nu^2)^2} \frac{\tanh(\pi\nu)}{\pi\nu} \frac{1/4+\nu^2}{1+\nu^2} \left[ \frac{s}{s_0} \right]^{\omega(\nu)}  \,\,.
\label{amplitudes_photo}\end{eqnarray}
\end{widetext}

The BFKL characteristic function   $\omega(\nu)$ is in general expressed as follows
\begin{eqnarray}
 \omega(\nu) = \overline{\alpha}_s \chi(\gamma) \label{eq:charac}
\end{eqnarray}
where  $\overline{\alpha}_s = (N_c \alpha_s)/ \pi$, $N_c$ is the number of  colors and   $\gamma = 1/2 + i\nu$. The function $\chi$ is given at leading order by   \cite{Lipatov:1976zz,Kuraev:1976ge,Kuraev:1977fs,Balitsky:1978ic}
\begin{equation}
 \chi^\mathrm{LO}(\gamma) =  2\psi(1) - \psi(\gamma) - \psi(1 - \gamma)
 \label{eq:orinlo}
\end{equation}
where $\psi(z)$ is the digamma function. This characteristic function has been used in Refs. \cite{Ivanov:1998jw,Cox:1999kv,Evanson:1999zb} in order to estimate the diffractive photon production. However, several shortcomings are present in a leading order calculation. Firstly,   the energy scale $s_0$ is arbitrary, which implies that the absolute value to the total cross section is therefore not predictable. Secondly, $\alpha_s$ is not running at LO BFKL. Finally, the power growth with energy violates $s$-channel unitarity at large rapidities. Consequently, new physical effects should modify the LO BFKL equation at very large $s$, making the resulting amplitude unitary.
 A theoretical possibility to modify this behaviour in a way consistent with the unitarity is the idea of parton saturation \cite{hdqcd}, where non-linear effects associated to high parton density are taken into account. Another possible solution, which is expected to diminishes the energy growth of the total cross section, is the calculation of higher order corrections to the BFKL equation. After an effort of several years, the next-to-leading order (NLO) corrections  were  obtained \cite{Fadin:1998py,Ciafaloni:1998gs}, with the  $\chi$ function being given by
\begin{equation}
 \chi(\gamma) = \chi^{\mathrm{LO}}(\gamma) + \overline{\alpha}_s \chi^{\mathrm{NLO}}(\gamma),
\end{equation}
where the correction term $ \chi^{\mathrm{NLO}}$ is given by \cite{Fadin:1998py,Ciafaloni:1998gs}
\begin{widetext}
\begin{eqnarray}
\chi^\mathrm{NLO} (\gamma) &=& {\cal C} \chi^\mathrm{LO} (\gamma) + \frac{1}{4}\left[\psi^{\prime\prime}(\gamma)+ \psi^{\prime\prime}(1-\gamma)\right] - \frac{1}{4}\left[\phi(\gamma)+ \phi(1-\gamma)\right] \nonumber \\
&& -\frac{\pi^2 \cos(\pi \gamma)}{4 \sin^2(\pi \gamma)(1-2 \gamma)}
\left\{3+\left(1+ \frac{N_f}{N_c^3}\right)\frac{(2+3\gamma(1-\gamma))}{(3-2\gamma)(1+2\gamma)}\right\} \nonumber \\
&& + \frac{3}{2} \zeta(3) - \frac{\beta_0}{8 N_c} \left(\chi^\mathrm{LO} (\gamma)\right)^2,  
\label{orignlo}
\end{eqnarray}
\end{widetext}
with ${\cal C} = \left(4-\pi^2 +5 \beta_0/N_c \right)/12$, $\beta_0 = (11N_c - 2N_f)/3$ is the leading coefficient of the QCD $\beta$ function, $N_f$ is the number of flavours, $\psi^{(n)}(z)$ is the poligamma function,  $\zeta(n)$ is the Riemann zeta-function and
\begin{widetext}
\begin{equation}
\phi (\gamma) + \phi (1-\gamma) = \sum_{m=0}^{\infty} \left[\frac{1}{\gamma+m}+\frac{1}{1-\gamma+m}\right]
\left[\psi^\prime(\frac{2+m}{2})-\psi^\prime(\frac{1+m}{2})\right].
\end{equation}
\end{widetext}
The  main problem associated to these NLO contributions is that they are so large that the problem appears perturbatively unstable. In particular, they imply negative corrections to the leading order kernel, resulting in a complex functional structure (pole positions) that gives an oscillating cross section \cite{ross}. Moreover,  there are also  problems associated to the choice of energy scale, the renormalization scheme and related ambiguities. 

An alternative to cure the highly unstable perturbative expansion of the BFKL kernel was proposed in Ref. \cite{Salam:1998tj}, who realized that the large NLO corrections emerge from the collinearly enhanced physical contributions. A method, the $\omega$-expansion, was then developed to resum collinear effects at all orders in a systematic way. The resulting  improved BFKL equation was consistent with  renormalization group requirements through matching to the DGLAP limit and resummation of spurious poles. In this approach the kernel is positive in a much larger region which includes the experimentally accessible one. This approach was revisited in Ref. \cite{Vera:2005jt} obtaining an expression for the collinearly improved BFKL kernel which does not mix longitudinal and transverse degrees of freedom and reproduces very closely the results from \cite{Salam:1998tj}. The   
characteristic function proposed  in Ref. \cite{Vera:2005jt}, denoted All-poles hereafter, is given by
\begin{widetext}
\begin{eqnarray}
\omega_{\mathrm{All-poles}} = \overline{\alpha}_s \chi^{\mathrm{LO}}(\gamma) + \overline{\alpha}_s^2 \chi^{\mathrm{NLO}}(\gamma) + \nonumber \\
+ \left\{\sum_{m=0}^{\infty}\left[\left(\sum_{n=0}^{\infty} \frac{(-1)^n(2n)!}{2^n n! (n+1)!} \frac{(\overline{\alpha}_s + a \overline{\alpha}_s^2)^{n+1} }{(\gamma+m-b\overline{\alpha}_s)^{2n+1}} \right)   - \frac{\overline{\alpha}_s}{\gamma+m} - \right.\right. \nonumber \\
\left. \left. -\overline{\alpha}_s^2\left(\frac{a}{\gamma+m} + \frac{b}{(\gamma+m)^2} - \frac{1}{2(\gamma + m)^3}\right) \right] + \{\gamma \rightarrow 1 - \gamma\}\right\} \,\,,
\label{eq:allpoles}
\end{eqnarray}
\end{widetext}
where 
\begin{eqnarray}
a = \frac{5\beta_0}{12 N_c} - \frac{13 N_f}{36N_c^3} - \frac{55}{36} \nonumber \\
b = - \frac{\beta_0}{8N_c} - \frac{N_f}{6N_c^3} - \frac{11}{12} \,\,.
\end{eqnarray}

Another alternative to solve the spurious singularities present in the original NLO kernel was proposed in Ref.  \cite{Brodsky:1998kn}. Differently from Refs. \cite{Fadin:1998py,Ciafaloni:1998gs}, where the calculations were performed by employing  the modified minimal subtraction scheme ($\overline{\mathrm{MS}}$) to regulate the ultraviolet divergences with arbitrary scale setting, in Ref. \cite{Brodsky:1998kn} they  propose to solve the energy scale ambiguity  using the Brodsky-Lepage-Mackenzie (BLM) optimal scale setting \cite{Brodsky:1982gc} and  the momentum space subtraction (MOM) scheme of renormalization.
In this approach, the  BFKL  characteristic function is given by
\begin{equation}
 \omega_\mathrm{BLM}^\mathrm{MOM} =  \chi^\mathrm{LO} (\gamma) \frac{\alpha_\mathrm{MOM}(\hat{Q}^{2}) N_c}{\pi} \left[1 + \hat{r}(\nu) \frac{\alpha_\mathrm{MOM}(\hat{Q}^{2})}{\pi} \right], \label{eq:omegablm}
\end{equation}
where $\alpha_\mathrm{MOM}$ is the coupling constant in the MOM scheme,
\begin{equation}
 \alpha_\mathrm{MOM} = \alpha_s\left[ 1 + \frac{\alpha_s}{\pi} T_\mathrm{MOM} \right]
\end{equation}
with $T$ being a function of number of colors, number of flavours and of a gauge parameter (See Ref. \cite{Brodsky:1998kn} for details) and
\begin{equation}
\alpha_s(\mu^2) = \frac{4\pi}{\beta_0 \ln (\mu^2/\Lambda^2_{\mathrm{QCD}})} \,\,. \label{eq:alfar}
\end{equation}
Moreover, the function $\hat{Q}$ is the BLM optimal scale, which is given by 
\begin{equation}
 \hat{Q}^2(\nu) = Q^2 \exp \left[ \frac{1}{2} \chi^\mathrm{LO}(\gamma) - \frac{5}{3} + 2 \left( 1 + \frac{2}{3}\varrho \right) \right], \label{eq:ascale}
\end{equation}
with $\varrho = -2\int^1_0 dx \ln(x)/(x^2-x+1) \approx 2.3439$.
 Finally, $\hat{r}$ is the NLO coefficient of the characteristic function,
\begin{widetext}
\begin{eqnarray}
 \hat{r}(\nu) &=&  -\frac{\beta_0}{4}\left[ \frac{\chi^\mathrm{LO}(\nu)}{2} -\frac{5}{3} \right] - \frac{N_c}{4\chi^\mathrm{LO}(\nu)} \left\{ \frac{\pi^2 \sinh(\pi\nu)}{2\nu \cosh^2(\pi\nu)} \left[ 3 + \left( 1 + \frac{N_f}{N_c^3} \right) \frac{11+12\nu^2}{16(1+\nu^2)} \right] \right.  \nonumber \\ 
&&   \left. - {\chi^{\prime\prime}}^\mathrm{LO}(\nu) + \frac{\pi^2-4}{3}\chi^\mathrm{LO}(\nu) - \frac{\pi^3}{\cosh (\pi \nu)} - 6\zeta(3) + 4\tilde{\phi}(\nu) \right\} + 7.471 - 1.281\beta_0 
\end{eqnarray}
\end{widetext}
with
\begin{equation}
 \tilde{\phi}(\nu) = 2 \int_0^1\!dx\,\frac{\cos(\nu\ln(x))}{(1+x)\sqrt{x}}\left[\frac{\pi^2}{6}-\mathrm{Li}_2(x)\right]\,\,,
\end{equation}
where $\mathrm{Li}_2(x)$ is the Euler dilogarithm or Spence function. 

\begin{figure}[t]
\begin{center}
 \includegraphics*[width=8cm]{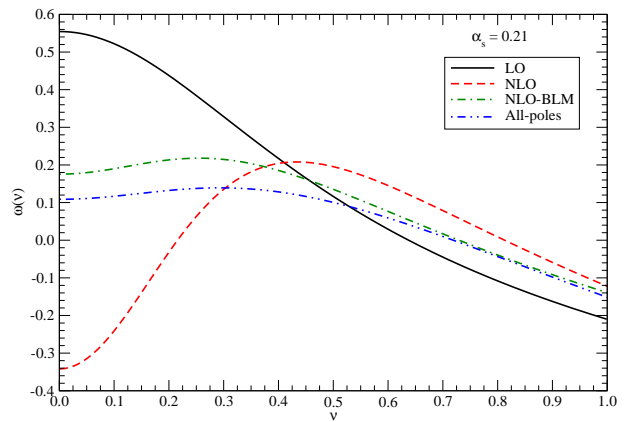}
\end{center}
\caption{ (color online). $\nu$ dependence of the characteristic function obtained considering NLO, NLO+All-poles and NLO+BLM approaches. For comparison, the behaviour of the LO BFKL characteristic function is also presented ($\alpha_s = 0.21$). }
 \label{fig:omega}
\end{figure}

In Fig. \ref{fig:omega} we present a comparison between the distinct  BFKL characteristic functions discussed above. For comparison we also present the LO BFKL function. We have that the All-poles  and BLM approaches regularize the behaviour of the original NLO BFKL characteristic function at small-$\nu$ and predict  similar behaviours for the function $\omega$. In particular, All-poles approach predicts that $\omega(0) = 0.11$ while BLM one predicts $\omega(0) = 0.18$. As we will show in the next section, this small difference have important implications in the energy dependence of the total cross section for diffractive photon production.

In the next section, we will use Eqs. (\ref{eq:orinlo}), (\ref{eq:allpoles}) and (\ref{eq:omegablm}) as input in  Eq.~(\ref{amplitudes}) in order to calculate the differential and total cross-section in the  BFKL formalism.

\section{Results \label{sec:res}}
In this section we present our results for the differential and total cross section and compare our predictions with the H1 experimental data \cite{Aaron:2008cq}. The differential cross section for the process $\gamma^\ast p \rightarrow \gamma X$ is given by 
\begin{equation}
 \frac{d\sigma}{d t} = \int^1_{x_\mathrm{min}} dx_j \frac{d\sigma (\gamma^\ast p \rightarrow \gamma X)}{dt dx_j}, \label{eq:dsdt}
\end{equation}
where we assume that $x_\mathrm{min} = 10^{-2}$  following Ref.  \cite{Aaron:2008cq} and previous theoretical studies  \cite{Forshaw:2001pf,Enberg:2002zy,Enberg:2003jw,Poludniowski:2003yk,Goncalves:2009gs,Goncalves:2010zb}. Moreover, the total cross section is obtained by integration of the Eq.    (\ref{eq:dsdt}) over the $|t|$ - range given by  $\unit[4.0]{GeV^2} < |t| < \unit[36.0]{GeV^2}$, according with H1 data \cite{Aaron:2008cq}. We will use the MSTW2008LO parametrization of the parton distribution functions \cite{Martin:2009iq}, but we have checked that a different parton parametrization modifies slightly the predictions.
In our calculations we have two free parameters: the strong coupling constant and the energy scale  parameter $s_0$. In what follows we assume that $\alpha_s = 0.21$, which is  the value used in Refs. 
\cite{Forshaw:2001pf,Enberg:2002zy,Enberg:2003jw,Poludniowski:2003yk,Goncalves:2009gs,Goncalves:2010zb}. It is important to emphasize that in our calculations $\alpha_s$ is fixed in the impact factors for all approaches studied, since they have been calculated at leading order, and running in the characteristic function when we are considering the All-poles and BLM approaches. In contrast with the vector meson production cross section, which is proportional to $\alpha \alpha_s^4$, in the photon production case, it is proportional to $\alpha^2\alpha_s^4$.  Following our previous studies \cite{Goncalves:2009gs,Goncalves:2010zb}, we have that $s_0 = c_1 \cdot Q^2 + c_2 \cdot |t|$. Considering that the H1 experimental data is for $Q^2 = 10^{-6}$ GeV$^2$ we will assume for simplicity that $Q^2 \approx 0$,  which implies $s_0 =  c_2 \cdot |t|$. Consequently, in our calculations we have only one free parameter: the coefficient $c_2$, which will be constrained by the H1 data for the differential cross section as in Refs. \cite{Goncalves:2009gs,Goncalves:2010zb}.

\begin{figure}[t]
\begin{center}
\includegraphics*[width=8cm]{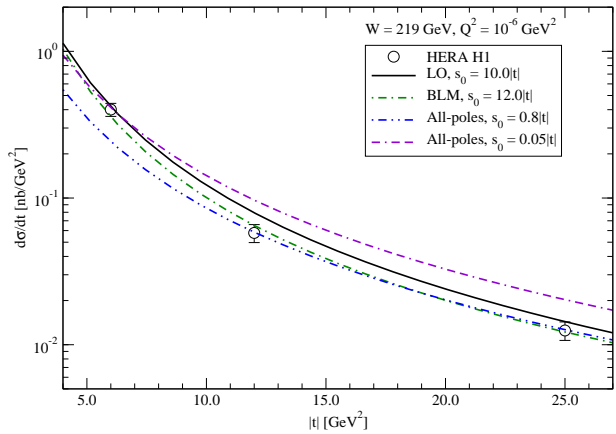} 
\end{center}
\caption{(color online). Differential cross section for the diffractive photon production at large-$t$ considering differential models for the characteristic function.  Experimental data from \cite{Aaron:2008cq}.}
\label{fig:odsdt}
\end{figure} 

In Fig. \ref{fig:odsdt} we present a comparison between our predictions for the differential cross section considering the different approaches for the characteristic function and the H1 data obtained for $Q^2 = 10^{-6}$ GeV$^2$ and $W = 219$ GeV. 
We adjust the value of $s_0$ in order to describe the larger number of experimental points. For the  LO BFKL approach we obtain that $c_2 = 10.0$ allow us to describe the data for low and high $|t|$. We have verified that if we adjust the value of $s_0$ in order to describe the data for $|t|$ = 12 GeV$^2$, the LO approach is not able to describe any other data point.
In contrast,  the value $c_2 = 12.0$ allow us to describe all data  using the BLM approach.  Using the All-poles approach we only describe the data for $|t| > $10 GeV$^2$ ($c_2 = 0.8$).  If we adjust $c_2$ in order to describe the data for $|t|$ = 6 GeV$^2$ ($c_2 = 0.05$) the approach does not describe the experimental data for larger values of $|t|$. Consequently, we can conclude that only the BLM approach allow us to describe all experimental data for the differential cross section. This conclusion also is valid if we consider the value of $\alpha_s$ as a free parameter. In this case the normalization is modified, which implies that a different of $c_2$ also should be used in order to describe the data, but we have verified the shape of the $|t|$-dependence is not strongly modified.

\begin{figure}[t] 
\begin{center}
\includegraphics*[width=8cm]{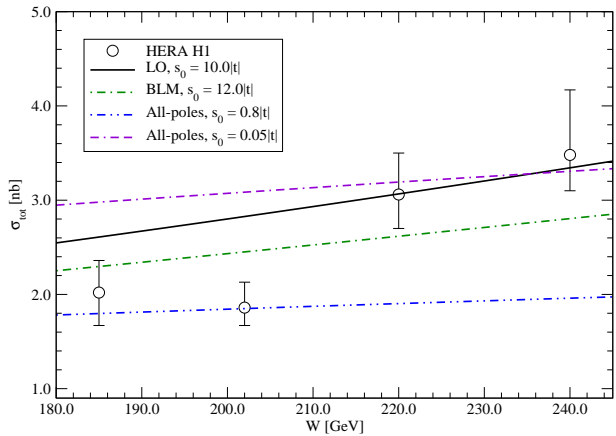} 
\end{center}
\caption{(color online). Total cross section for the diffractive photon production at large-$t$ considering differential models for the characteristic function.  Experimental data from \cite{Aaron:2008cq}.}
\label{fig:ostot}
\end{figure}

Having fixed the free parameter in our calculations, we now are able to predict the energy dependence of the total cross section considering the different approaches for the BFKL characteristic function. In Fig. \ref{fig:ostot} we present our predictions. We can see that the LO BFKL approach predicts, as expected, the steeper energy growth of the total cross section. In contrast, the All-poles approach predicts the milder energy dependence. In comparison with the experimental data, anyone of the approaches describe all experimental data for the total cross section. The BLM approach, which describes quite well the differential cross section,  is able to describe the data at low energy but underestimate the total cross section  at high energies. The All-poles prediction is strongly dependent on the parameter $c_2$. If  $c_2 = 0.8$ is used, this approach describes the data for low energies but underestimate the total cross section at high energies. In contrast, at $c_2 = 0.05$, this approach is only able to describe the data for $W> 210$ GeV.

Some comments are in order before we summarize our main conclusions. First, the current experimental data for the diffractive photon production are scarce and have been obtained in a limited energy range. Certainly, more data and for larger values of energy should allow us to obtain more definitive conclusions about the underlying QCD dynamics.  
The second, and more important, comment is related to the fact that in our calculations we are using leading order impact factors. In a full NLO calculation  one should also consider the NLO corrections to the impact factors. In the last few years, the real and virtual corrections which contributes at NLO has been estimated \cite{Fadin:2002tu,Bartels:2000gt,Bartels:2001mv,Bartels:2002uz,Bartels:2004bi} and  recently analytical expressions have been proposed \cite{Balitsky:2012bs}.    Numerical results were presented in Refs. \cite{Bartels:2004bi,grigorios} and indicate that the NLO corrections  tend to decrease the value of the impact factors.  Consequently,  our estimates for the total cross section using the NLO-BFKL characteristic functions should be consider an upper bound. It is important to emphasize that the procedure used in our analysis was  used in Refs. \cite{Ivanov:2005gn,Ivanov:2006gt,Enberg:2005eq} in order to estimate the  
the double production of vector mesons ($\gamma^{\ast}\gamma^{\ast} \rightarrow VV$) and in Refs. \cite{Goncalves:2006mg,Caporale:2008is} to calculate the  $\gamma^{\ast}\gamma^{\ast}$ total cross section.
Moreover, in Refs. \cite{Vera:2006un,Vera:2007kn,Marquet:2007xx}, the production of  Mueller-Navelet jets was investigated considering the NLO BFKL kernel and LO impact factors and, more recently, the description of the $F_2$ structure function  has been performed using a similar approach \cite{Hentschinski:2012kr,Hentschinski:2013id} \footnote{A full NLO BFKL analysis of the Mueller-Navelet jet production was presented in Ref. \cite{wallon}. }. All these studies demonstrate that the study of the diffractive photon production modifying only the BFKL kernel is justified in order to a first estimate of the magnitude of the NLO contributions for this process.

\section{Summary \label{sec:concl}}

The description of the 
high energy limit of the Quantum Chromodynamics (QCD) is an important open question in the Standard Model. During the last decades several approaches were developed in order to improve our understanding from a fundamental perspective. In particular, 
after a huge theoretical effort, now we have available the NLO corrections for the BFKL characteristic function, which allow us to improve the analysis of the processes which are expected to probe the underlying QCD dynamics.  In this paper we have studied, for the first time, the diffractive photon production considering the NLO characteristic functions proposed in Refs. \cite{Vera:2005jt,Brodsky:1998kn}. Moreover, the LO BFKL approach for the process was revisited. The free parameters in our calculations has been constrained considering the H1 data for the differential cross section at fixed energy and after we have predicted the energy dependence of the total cross section. We obtain a reasonable description of the scarce experimental data.   New data in a large energy range should allow us to obtain more definitive conclusions about the QCD dynamics. Moreover, the inclusion of the NLO correction in the impact factor is an important step which we expect to  perform in a next future.

\section*{Acknowledgements}
This work was partially financed by the Brazilian funding agencies CNPq and CAPES.


\end{document}